\documentclass[preprint]{aastex63}       %one column except with 12 point font.

\usepackage{longtable} % for 'longtable' environment
\usepackage{threeparttable} % for "ThreePartTable" environment
\usepackage{booktabs}    %to use \toprule, \midrule, and \bottomrule
\usepackage{lineno}
\usepackage{bibunits}
\usepackage{bm}
\usepackage{amsmath, amssymb}
\usepackage{rotating}

\accepted{by ApJ}
\shortauthors{Hou et al.}
\begin{document}
\title{Monte-Carlo simulations on possible collimation effects of outflows to fan-beamed emission of ultraluminous accreting X-ray pulsars} 
%\title{The three musketeers share similar emission properties with bright Galactic and extragalactic accreting X-ray pulsars} %\textcolor{blue}{may need change}}

\correspondingauthor{X. Hou, Y. You, S.N. Zhang}
\email{xhou@ynao.ac.cn, youyuan@ihep.ac.cn, zhangsn@ihep.ac.cn}

\author[0000-0003-0933-6101]{X. Hou}
\affiliation{Yunnan Observatories, Chinese Academy of Sciences, Kunming 650216, China}
\affiliation{Key Laboratory for the Structure and Evolution of Celestial Objects, Chinese Academy of Sciences, Kunming 650216, China}

\author[0000-0002-0352-8148]{Y. You}
\affiliation{Key Laboratory for Particle Astrophysics, Institute of High Energy Physics, Chinese Academy of Sciences, Beijing 100049, China}
\affiliation{University of Chinese Academy of Sciences, Chinese Academy of Sciences, Beijing 100049, China}

\author[0000-0001-9599-7285]{L. Ji}
\affiliation{School of Physics and Astronomy, Sun Yat-Sen University, Zhuhai 519082, China}

\author[0000-0002-4622-796X]{R. Soria}
\affiliation{University of Chinese Academy of Sciences, Chinese Academy of Sciences, Beijing 100049, China}
\affiliation{Sydney Institute for Astronomy, School of Physics A28, The University of Sydney, Sydney, NSW 2006}

\author[0000-0001-5586-1017]{S.N. Zhang}
\affiliation{Key Laboratory for Particle Astrophysics, Institute of High Energy Physics, Chinese Academy of Sciences, Beijing 100049, China}
\affiliation{University of Chinese Academy of Sciences, Chinese Academy of Sciences, Beijing 100049, China}

\author[0000-0002-2749-6638]{M.Y. Ge}
\affiliation{Key Laboratory for Particle Astrophysics, Institute of High Energy Physics, Chinese Academy of Sciences, Beijing 100049, China}

\author[0000-0002-2705-4338]{L. Tao}
\affiliation{Key Laboratory for Particle Astrophysics, Institute of High Energy Physics, Chinese Academy of Sciences, Beijing 100049, China}

\author{S. Zhang}
\affiliation{Key Laboratory for Particle Astrophysics, Institute of High Energy Physics, Chinese Academy of Sciences, Beijing 100049, China}

\author[0000-0001-7584-6236]{H. Feng}
\affiliation{Department of Astronomy, Tsinghua University, Beijing 100084, China}

\author{M. Zhou}
\affiliation{Yunnan Observatories, Chinese Academy of Sciences, Kunming 650216, China}
\affiliation{Key Laboratory for the Structure and Evolution of Celestial Objects, Chinese Academy of Sciences, Kunming 650216, China}

\author[0000-0003-3127-0110]{Y.L. Tuo}
\affiliation{Key Laboratory for Particle Astrophysics, Institute of High Energy Physics, Chinese Academy of Sciences, Beijing 100049, China}

\author[0000-0003-0274-3396]{L.M. Song}
\affiliation{Key Laboratory for Particle Astrophysics, Institute of High Energy Physics, Chinese Academy of Sciences, Beijing 100049, China}
\affiliation{University of Chinese Academy of Sciences, Chinese Academy of Sciences, Beijing 100049, China}

\author{J.C. Wang}
\affiliation{Yunnan Observatories, Chinese Academy of Sciences, Kunming 650216, China}
\affiliation{Key Laboratory for the Structure and Evolution of Celestial Objects, Chinese Academy of Sciences, Kunming 650216, China}
\affiliation{University of Chinese Academy of Sciences, Chinese Academy of Sciences, Beijing 100049, China}

%% Mark off the abstract in the ``abstract'' environment. 
\begin{abstract}

Pulsating ultraluminous X-ray sources (PULXs) are accreting pulsars with apparent X-ray luminosity exceeding $10^{39}\, \rm erg\ s^{-1}$. We perform Monte-Carlo simulations to investigate whether high collimation effect (or strong beaming effect) is dominant in the presence of accretion outflows, for the fan beam emission of the accretion column of the neutron stars in PULXs. We show that the three nearby PULXs (RX J0209.6$-$7427, Swift J0243.6+6124 and SMC X-3), namely the three musketeers here, have their main pulsed emission not strongly collimated even if strong outflows exist. This conclusion can be extended to the current sample of extragalactic PULXs, if accretion outflows are commonly produced from them. This means that the observed high luminosity of PULXs is indeed intrinsic, which can be used to infer the existence of very strong surface magnetic fields of $\sim10^{13-14}$ G, possibly multipole fields. However, if strong outflows are launched from the accretion disks in PULXs as a consequence of disk spherization by radiation pressure, regular dipole magnetic fields of $\sim10^{12}$ G may be required, comparable to that of the three musketeers, which have experienced large luminosity changes from well below their Eddington limit ($2\times10^{38}\, \rm erg\ s^{-1}$ for a NS) to super-Eddington and their maximum luminosity fills the luminosity gap between Galactic pulsars and extragalactic PULXs. 
%Our results suggest that 
%all these sources belong to the same class of accreting pulsars, and 
%the high luminosity of extragalactic PULXs only samples the very rare bright accreting X-ray pulsars in those galaxies.
%We also find that the three musketeers and extragalactic PULXs share the same range of surface magnetic fields ($\sim 10^{12-13}$ G) estimated with the commonly used Ghosh \& Lamb torque model, which is also similar to that measured from the detected cyclotron resonant scattering features (CRSFs) of most Galactic accreting X-ray pulsars. Our results suggest that all these sources belong to the same class of accreting pulsars, and the high luminosity of extragalactic PULXs only samples the very rare bright accreting X-ray pulsars in those galaxies.

\end{abstract}

%% Keywords should appear after the \end{abstract} command. 
%% See the online documentation for the full list of available subject
%% keywords and the rules for their use.

\section{Introduction} \label{sec:intro}

Ultraluminous X-ray sources (ULXs) are off-nucleus objects whose apparent X-ray luminosity exceeds $10^{39}\, \rm erg\ s^{-1}$, the Eddington limit of a typical stellar mass black hole ($\sim 10\rm M_{\odot}$) \citep{Kaaret2017}. 
%Initially, two main classes of models have been proposed to explain their super-Eddington luminosity: one is intermediate-mass BHs ($10^{2}-10^{5} \,M_{\rm \odot}$) with sub-Eddington accretion  \citep{Colbert1999,Makishima2000,Kaaret2001,Miller2003,Feng2011,Sutton2012}, the other is normal stellar-mass X-ray binaries accreting at or above the Eddington limit  \citep{King2001,Gao2003,Gilfanov2004,Roberts2004,Stobbart2006,Poutanen2007,Roberts2007,Zampieri2009,Feng2011}. 
Recent studies revealed that some of them are accreting neutron stars (NSs), called pulsating ULXs (PULXs), the luminosity of which is usually more than $\sim2-3$ orders of magnitude higher than most Galactic and nearby accreting X-ray pulsars, including those in the Large Magellanic Cloud (LMC) and the Small Magellanic Cloud (SMC). 

It has been suggested that the NSs in PULXs have very high magnetic fields, higher than that of the regular accreting NSs ($\sim 10^{12}$ G), or even comparable to that of magnetars ($\sim10^{14} $ G) \citep{Mushtukov2015b,Chashkina2017,Chashkina2019,Israel2017a,Eksi2015,DallOsso2015,Tsygankov2016b,Tsygankov2017,Brice2021}, in order to explain their extremely high luminosity. Currently, multipole magnetic fields near the NS surfaces have been proposed for at least five PULXs, including Swift J0243.6+6124 \citep{Kong2022}, which is the first Galactic PULX, RX J0209.6$-$7427 \citep{Hou2022} and SMC X-3 \citep{Tsygankov2017} which are in the SMC, M82 X-2 \citep{Eksi2015} and NGC 5907 ULX-1 \citep{Israel2017a} which are extragalactic sources. Especially, the detection of the cyclotron resonance scattering feature (CRSF) in Swift J0243.6+6124, which has the highest CRSF energy to date among all X-ray pulsars, unambiguously proves the presence of multipole field ($\gtrsim 1.6\times 10^{13}$~G) close to the surface of the NS, since its dipole field inferred from the accretion disk-magnetosphere interaction is much weaker \citep{Kong2022}. On the other hand, as shown in \citet{Erkut2020} for most of the PULXs (but without including RX J0209.6$-$7427), the ranges of dipole magnetic fields (by implicitly assuming that the possible multipole fields are not important) obtained by taking into account different assumptions on the states of interactions between the NS and accretion disk, as well as by assuming the observed peak luminosity as the maximally allowed luminosity, can cover the field strength inferred from the CRSF of Swift J0243.6+6124. Nevertheless, 
% due to the still relatively small sample of PULXs, (given the observed NS spin and luminosity)
it remains unclear whether very high magnetic fields are necessary to explain the ultraluminous emission of PULXs and whether multipole magnetic fields are common in PULXs. 
 
Some other models of PULXs include strongly collimated (or beamed) radiation\footnote{Here we prefer ``collimated" to ``beamed", since pencil ``beam" and fan ``beam" are the standard terminologies to describe the emission patterns of accreting X-ray pulsars. However, the directly observed emission of a fan ``beam" is close to the intrinsic luminosity of a pulsar, and thus not amplified; a pencil ``beam" may be confined to a certain solid angle depending on the emission profile of the ``pencil". On the other hand, ``beamed" emission usually refers to strongly amplified emission in a small solid angle, which we refer to ``collimated" emission, since the amplification process is assumed to be due to the ``collimation" effect of the outflows.}, which is confined within a small solid angle and towards the observer, such that the intrinsic luminosity is much lower than that directly inferred from the observed X-ray flux \citep{King2009,Kluzniak2015,King2016,Koliopanos2017,Pintore2017,King2017,Middleton2017,Walton2018,King2019,King2020}. The recent theoretical work \citep{Erkut2020} systematically studied the magnetic fields and collimation effect of all known PULXs at that time with alternative assumptions and showed no bias in favor of the collimated radiation or the existence of magnetar fields. They argued that PULXs do not require magnetar-strength surface dipole fields if collimation is taken into account, yet the fields are strong enough for the NSs in ULXs to magnetically channel the accretion flow in supercritical (super-Eddington) accretion disks.

Our previous study \citep{Hou2022} has demonstrated that the main pulsed emission of RX J0209.6$-$7427 is from the ``fan beam'' pattern which is fundamentally not collimated to a specific direction, consistent with the observed sinusoidal pulse profiles of the current PULX sample \citep{Kaaret2017}.
The recent study of Swift J0243.6+6124 during its 2020 super-Eddington outburst by \cite{Liu2022} revealed that its emission is also fan beamed basing on the pulse profile evolution. During the 2016-2017 super-Eddington outburst of SMC X-3 \citep{Townsend2017,Weng2017,Tsygankov2017,Zhao2018}, its pulse profile in the low energy range of 0.5$-$10 keV has a complex three-peak structure \citep{Koliopanos2018}. The authors surmised that the softer and more prominent peak is from the fan beam, while the two harder peaks are from the ``polar beam'' emission which is produced by reflection of the incident fan beam photons off the surface of the NS.
Therefore, the ultraluminous emission of the three close PULXs, i.e., the three musketeers, are all fan beam in origin, thus not strongly collimated.

%exhibits more complex profile evolution with luminosity \citep{Doroshenko2020} than RX J0209.6$-$7427. The pulse profile evolution of Swift J0243.6+6124 with energy around the highest luminosity is very similar to RX J0209.6$-$7427, the minor peak getting less prominent at higher energies; the highest energy bin with significant pulsation detection is 50$-$100 keV \citep{Wang2020}.  

%SMC X-3 is another close PULX in the SMC. During its 2016-2017 super-Eddington outburst \citep{Townsend2017}, its pulse profile in the low energy range of  0.5$-$10 keV has a complex three-peak structure \citep{Koliopanos2018}. Basing on a simple toy model, the authors surmised that the softer and more prominent peak is from the fan beam, while the two harder peaks are from the ``polar beam'' emission which is produced by reflection of the incident fan beam photons off the surface of the NS. However, \textit{NuSTAR} observations revealed inconsistent phenomenon \citep{Tsygankov2017a}: the softer peak is only visible in 3$-$10 keV while the two harder peaks are prominent in 40$-$79 keV. This suggests instead that the two harder peaks should be preferentially from the fan beam, since in the simple toy model the fan beam emission should dominate such high energy photons. Again, as in the case of Swift J0243.6+6124, the lack of the knowledge of higher energy pulsations prevents us from a conclusive determination of the emission origin. Such firm determination can only be achieved with RX J0209.6$-$7427 owing to its detected pulsation above 130 keV with \insight{}. 

Extragalactic PULXs are also in super-Eddington accretion although only one main pulse peak is detected so far. It is therefore reasonable to extend our conclusion, i.e., the main emission of the three musketeers is from the fan beam without strong collimation, to extragalactic PULXs. This is because the higher the accretion rate, the more likely the radiation from the AC can escape in the form of a fan beam. Other studies also show that Swift J0243.6+6124, SMC X-3 and NGC 300 ULX1 only have small or moderate collimation \citep{King2020,Binder2018}, which is consistent with our argument. However, these studies were all based on the BH outflow theory %assumed pencil-like pulsed emission \citep[i.e., ``beaming factor"$<1$ in][]{King2020,Binder2018}, 
and did not consider the accretion column (AC) physics. Besides, the high spin frequency derivative $\nu_{1}\sim 10^{-10}\ \rm Hz\, s^{-1}$ observed in the three musketeers is consistent with that of extragalactic PULXs \citep{King2020,Fabrika2021}, implying a high accretion rate and thus a high intrinsic luminosity. This supports as well that strong collimation is unnecessary to explain the super-Eddington luminosity of PULXs. 

However, high degree of collimation has been suggested, especially if strong outflows are produced at very high accretion rate \citep{King2009,Kluzniak2015,King2016,Koliopanos2017,Pintore2017,King2017,Middleton2017,Walton2018,King2019,King2020}. This is one of the main models explaining the super high luminosity of extragalactic PULXs. Strong collimation requires sufficient reprocessing in the outflow which will soften the energy spectrum, and consequently smear or broaden the pulse peak and effectively reduce the pulse fraction (PF) at low energies. This is, however, not observed in RX J0209.6$-$7427. Its main peak remains from 1 keV to above 130 keV and the PF is much larger below 50 keV than at higher energies \citep[Figure 7 in][]{Hou2022}). This suggests that the main radiation is observed directly without reprocessing, thus providing additional evidence that the main pulsed emission of RX J0209.6$-$7427 is not strongly collimated.  
%The rapid decrease of PF with energy at above 50 keV might be due to the reflection by the NS's surface, which will smear/broaden the pulse peak and thus reduce the PF, if these high energy photons are emitted in fan beam near the bottom of the AC, as illustrated in Figure~\ref{fig:geometry}. Xian: this is has been discussed in the first ApJ paper.

Recently, simulations of PULXs with the presence of accretion outflow show that the large PF detected in PULXs excludes large luminosity amplification caused by the geometrical collimation due to outflows, and only a negligible fraction of strongly collimated PULXs can show PF above 10\% \citep{Mushtukov2021}. The PFs of RX J0209.6$-$7427 \citep[15\%$-$55\%, see Figure 7 in][]{Hou2022}, Swift J0243.6+6124 \citep[20\%$-$85\%,][]{Tao2019} and SMC X-3 \citep[30\%$-$100\%,][]{Weng2017,Tsygankov2017} during the outbursts are all very high, consistent with non or low collimation. Similarly, extragalactic PULXs also have sinusoidal profiles and high PF, thus are impossible to be strongly collimated based on the simulation result. 

However, these simulations were based on the pencil beam emission pattern of the AC \citep{Mushtukov2021}, which we have proven very unlikely for the ultraluminous emission of PULXs \citep{Hou2022}. Therefore, we perform in this work simulations for the fan beam pattern to investigate the collimation effect due to outflows (Section \ref{MC}). We find that the collimation effect is even less significant, and in most cases the outflows will actually block the fan beam emission of the AC. We discuss our results in Section \ref{discuss}.
%including further links with extragalactic PULXs of the three sources, and the comparison of the Galactic accreting X-ray pulsars with extragalactic PULXs on the footing that their luminosity is intrinsic rather than highly collimated. 

%and found that the collimation factor of outflow would be even more reduced. Under the fan beam pattern, in most cases the collimation factor is less than 3, and can reach about one order of magnitude only in extreme cases.  All these results robustly demonstrate that the main radiations of these sources are not strongly collimated and their apparent luminosity is close to that emitted from the AC. 

\section{Monte Carlo simulation of collimation effect}
\label{MC}
We use a Monte Carlo method to trace the scattering history of X-ray photons emitted from the AC of an accreting NS. The purpose of the simulation is to study whether a presumed outflow will cause a strong collimation effect on the radiation of an accreting NS. Different from the previous simulations \citep{Mushtukov2021}, the collimation effects to the fan beam emission of the AC are simulated here. 

\subsection{Geometry and Methodology}

\begin{figure}[ht]
    \centering
    \includegraphics[scale=0.3]{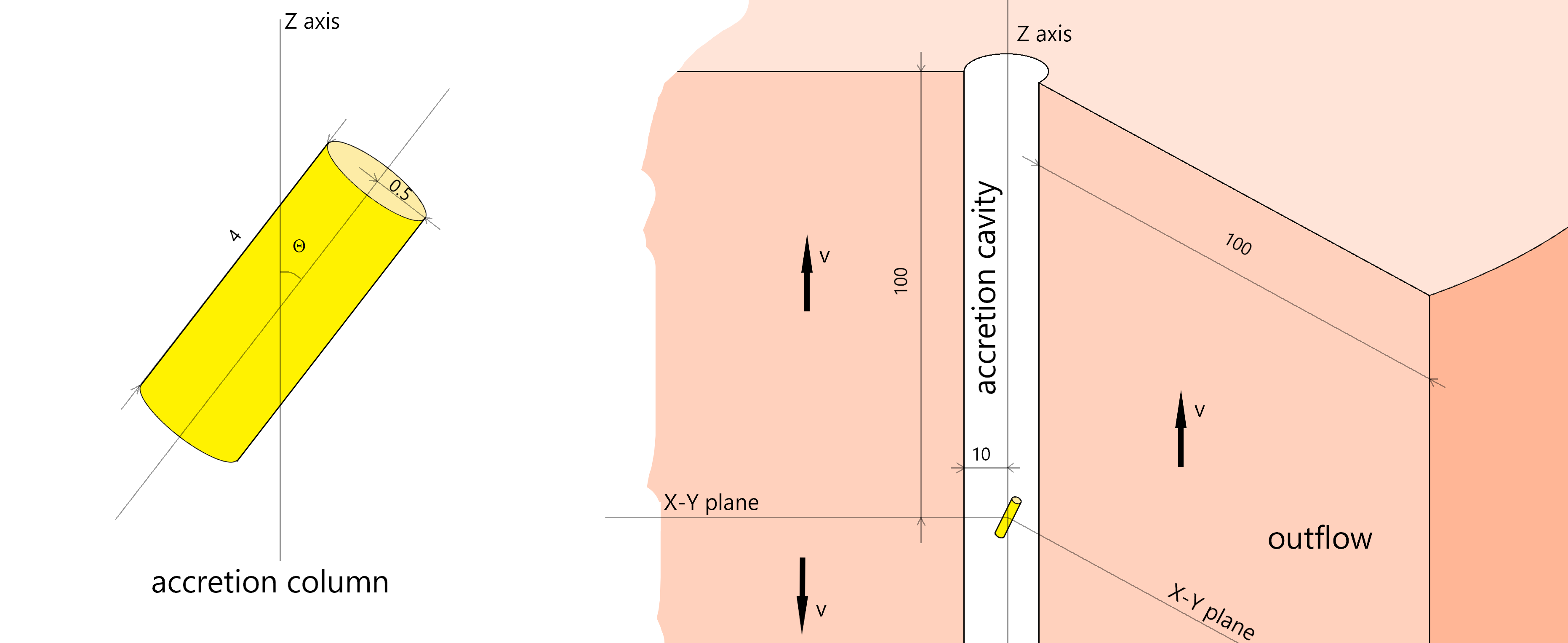}
    \caption{ \footnotesize Schematic illustration of the simulated geometry. The AC of the NS emits seed photons at the center of the accretion cavity, surrounded by an axially symmetric outflow.}
    \label{fig:simulation_illustration}
\end{figure}

Assuming that the NS is located within the wall of an accretion cavity; the outflow is beyond the wall which consists of vertically out-flowing thermal electrons with uniform density. Figure \ref{fig:simulation_illustration} shows the schematic illustration of this geometry, which is similar to that in  \cite{Mushtukov2021}. Unlike the simulation in \cite{Mushtukov2021}, the electrons in our simulation are not static and the outflow has finite optical depth which is 100 in both vertical and radial directions. The assumed optical depth of 100 is to mimic the ``infinite" optical depth assumed in \cite{Mushtukov2021} without causing too much computational burden. Indeed, for a test case we find that an optical depth of 500 results in qualitatively similar results. In the outflow, the density of electrons is set to be unity, thus the Thomson optical depth of the outflow is equal to the geometrical distance. We set the ratio between the height and the radius of the accretion cavity as 10, thus the radius of the accretion cavity is 10. For a smaller ratio, the collimation factor can be increased, but the fraction of NSs blocked by the cavity is higher to the line of sight of the observer; therefore the exact choice of the ratio does not change our conclusions. The ratio of 10 is chosen simply as a reasonable trade-off between the expected collimation factor and fraction of the NSs blocked by the cavity. The motion of the electrons is a combination of random motion with 1 keV kinetic energy and vertical bulk motion with velocity $v$. The 1 keV kinetic energy is used to mimic the thermal X-ray emitting winds possibly observed in some accreting X-ray binary systems; the exact choice of the temperature in the outflow does not change our conclusions, as far as the temperature is significantly lower than the break energy of the power-law, which is around 10 keV.

As shown in \cite{Hou2022}, we have demonstrated that the ultraluminous X-ray radiation of the AC should be fan beam. We ignore the size of the NS and regard the AC of the NS as a cylinder. This cylinder is the only source of initial photons in the system. The height and radius of the AC cylinder are set as 4 and 0.5 in units of optical depth, to make the AC as a thin cylinder but with a size much smaller than the cavity. We have tested that the exact size of AC will not make significant difference on the simulation results, but the ratio between the height and radius of the accretion cavity is important. The angle between the axis of the AC and the axis of the accretion cavity is $\Theta$. The NS does not spin in our model, although NSs are generally spinning. Our simulated radiation can thus be considered as phase-averaged radiation for a spinning NS. The seed photons are emitted from the side face of the AC cylinder, with an assumed angular distribution:
\begin{equation}\label{equation:seed}
    P_{\rm seed} \varpropto \cos\theta,
\end{equation}
where $\theta \in [0,\pi /2]$ is the angle between the normal vector of the side face of the AC and the momentum of the seed photon, $P_{\rm seed}$ is the emitting probability of the seed photon. In this way, the luminosity of the AC in one direction is proportional to the projected area of the side face of the AC in this direction. The energy of seed photons is assumed to follow the cutoffpl model in XSPEC, with power law photon index 0.5 and e-folding energy (exponential rolloff) 8.1 keV, which is the average spectrum of known PULXs \citep{Walton2018}.

\begin{figure}[ht]
    \centering
    \includegraphics[scale=0.16]{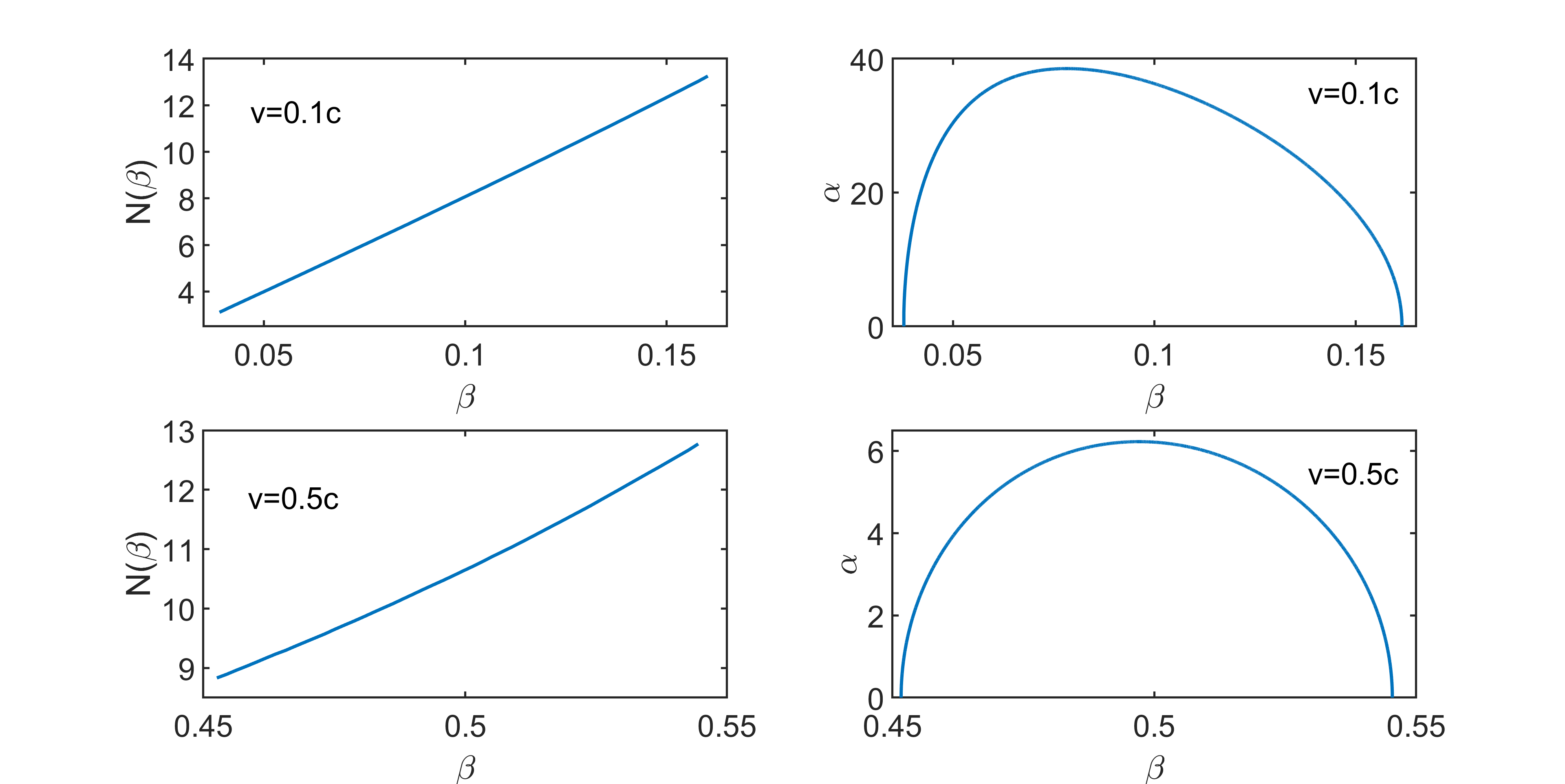}
    \caption{ \footnotesize The combination of thermal motion and bulk motion of the electrons. The left two panels show the probability density distribution functions of the velocity of electrons $N(\beta)$, with bulk velocity $v=0.1c$ and $v=0.5c$ respectively. The right two panels show the relationship between the direction of motion of an electron and the velocity of the electron with $v=0.1c$ and $v=0.5c$, respectively, in which $\alpha$ is the angle between the axis of the cavity and the direction of motion of the electron.}
    \label{fig:simulation_electron}
\end{figure}

We make three sets of simulations to discuss the collimation effects of the outflow with velocity $v$ and of the inclination angle $\Theta$ of the AC, which are $\Theta = 0$ and $v=0.1c$, $\Theta = 45^{\circ}$ and $v=0.1c$, $\Theta = 0$ and $v=0.5c$. Generally we follow the calculation method of Compton scattering in \cite{Pozdnyakov1983} (see equations in Section 2.1), and the steps of the Monte Carlo simulations are as follows:

(I)	Calculate the cross section of Compton scattering. The total cross section of Compton scattering is:
\begin{equation}
\begin{aligned}
    \sigma=&(1-{\bf u}\cdot{\bm\omega}/c)\cdot\frac{3}{4x}\sigma_{\rm T}\bigg{[}(1-\frac{4}{x}-\frac{8}{x^2})\ln(x+1)\\&+\frac{1}{2}+\frac{8}{x}-\frac{1}{2(x+1)^2}\bigg{]},
\end{aligned}
\end{equation}
with
\begin{equation}
    x=\frac{2h\nu}{m_ec^2}\gamma(1-{\bf u}\cdot{\bm\omega}/c),
\end{equation}
where $\nu$ and ${\bm\omega}$ are the frequency and the direction of the photon, ${\bf u}$ is the velocity of the electron, $\gamma$ is its Lorentz factor, $x$ is the photon energy in the rest frame of the electron (in units of $m_ec^2$) and $\sigma_{\rm T}$ is the Thomson cross section. In the outflow, electrons have 1 keV energy and move isotropically. Therefore, the average total cross section can be calculated as
\begin{equation}
    \bar{\sigma}(\nu,{\bm\omega})=\frac{\int\sigma(x,\beta)N(\beta)d\beta}{\int N(\beta)d\beta},
\end{equation}
where $\beta$ is the velocity of the electron in units of $c$, and $N(\beta)$ is the probability density distribution function of $\beta$ due to the combination of thermal motion and bulk motion of the electrons, as shown in Figure \ref{fig:simulation_electron}. In this step, we calculate $\bar{\sigma}(\nu,{\bm \omega})$ at different $\nu$ and ${\bm\omega}$ in advance to form a database, so that $\bar{\sigma}(\nu,{\bm\omega})$ at any $\nu$ and $\bm\omega$ needed in the future steps can be calculated by an interpolation method.

(II) Generate a seed photon. When sampling the emitted location of the seed photons, the probability is assumed the same everywhere on the side surface of the AC. The momentum of a seed photon follows the laws described above. Each set of simulation contains $10^7$ photons emitted from the AC.

(III) A photon not in the outflow moves along straight trajectory until it reaches the wall of the accretion cavity, or enters the AC, or escapes from the accreting NS system. For the latter two cases, the simulation stops and turns back to step (II).

(IV) For a photon located in the outflow, Compton scattering will be simulated. For a geometrical distance $l$ from the current location of the photon to the edge of the outflow along the moving direction of the photon, the probability of the photon escaping from the outflow is
\begin{equation}
    P_{\rm escape} = \exp(-\frac{\bar{\sigma}(\nu,{\bm\omega})}{\sigma_{\rm T}}l).
\end{equation}
If the photon escapes, we change the location of the photon to the edge of the outflow and keep the momentum of the photon unchanged; the simulation then returns to step (III). If the photon fails to escape from the outflow, then the free path of the photon before the next scattering can be obtained by sampling 
\begin{equation}
    \delta l = -\frac{\sigma_{\rm T}}{\bar{\sigma}(\nu,{\bm\omega})} \cdot \ln(1-X \cdot P_{\rm escape}),
\end{equation}
where $X \in [0,1]$ is a random number. With the scattering position determined, the scattering will be simulated in the next step.

(V) To simulate a scattering, first we need to sample the velocity of the electron ${\bf u}$ involved in this scattering. For a given photon $(\nu, \bm\omega)$, the probability of this photon being scattered by an electron with velocity $\bf u$ is proportional to $\sigma \cdot N(\beta)$. The momentum of the scattered photon is determined by the differential cross section of the scattering. In order to simplify the calculation, we transform the reference frame from the rest frame of the NS to the rest frame of the electron. In this reference frame, the differential cross section has a simple form as
\begin{equation}\label{equation:differential}
    \frac{d\sigma}{d\omega}=q^2(q+\frac{1}{q}-\sin^2\eta),
\end{equation}
where $q$ is the energy ratio between the scattered photon and the incident photon, and $\eta$ is the scattering angle. We sample $q$ and $\eta$ according to equation (\ref{equation:differential}), transform them back to the NS's reference frame and get the momentum of the photon after the scattering. Now the simulation returns to step (IV) with the location and momentum of the scattered photon.

\subsection{Results of simulations}

\begin{figure}[ht]
    \centering
    \includegraphics[scale=0.3]{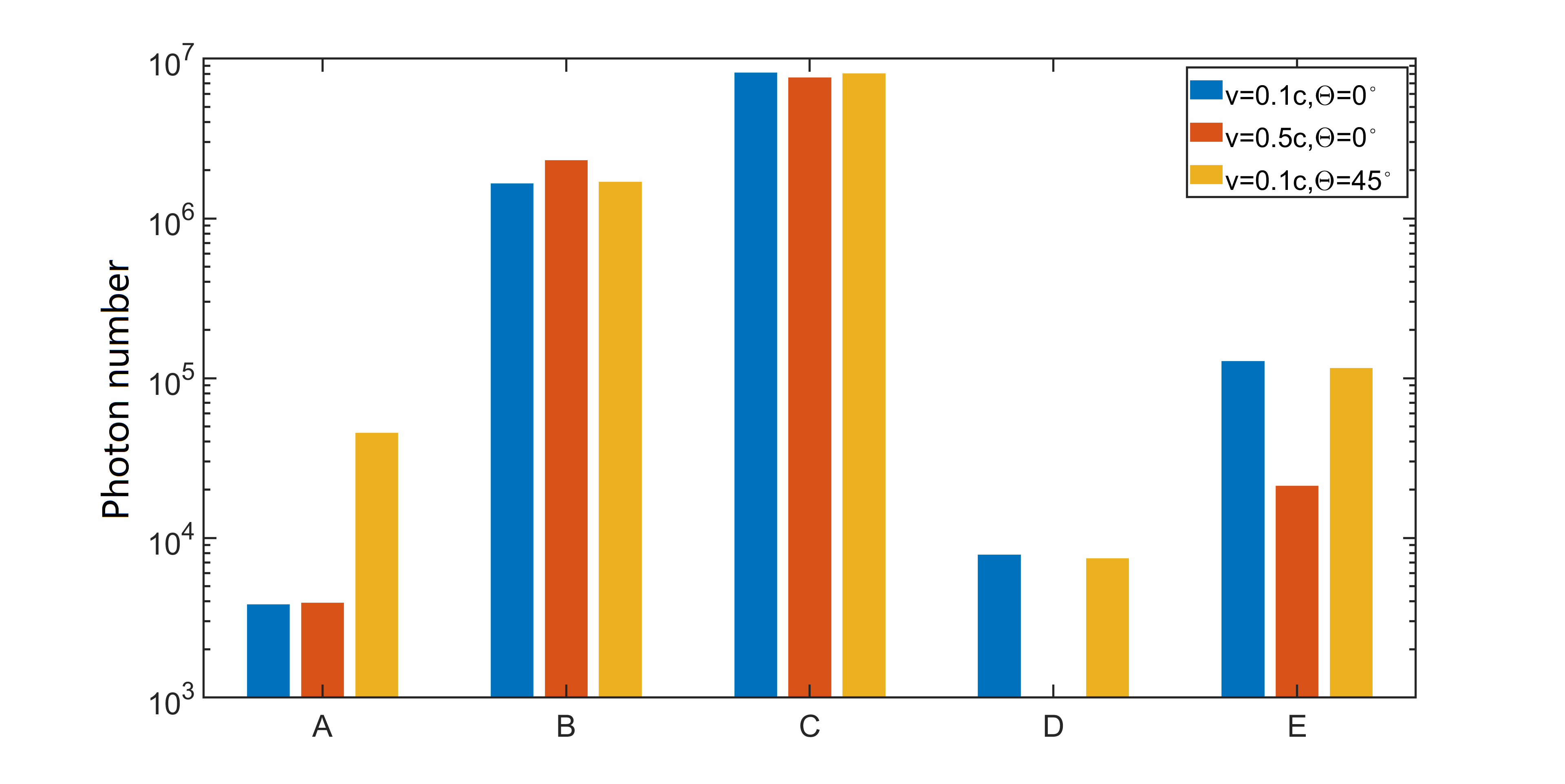}
    \caption{ \footnotesize Photons classified by their escaping places from the system. ``A": seed photons that directly leave the accretion cavity without any interaction. ``B": scattered photons that escape the system from the wall of the cavity. ``C": scattered photons that pass through the top/bottom surfaces of the outflow. ``D": scattered photons that leave the system from the side surface of the outflow. ``E": photons that escape from the wall of the cavity and entered into the AC.}
    \label{fig:simulation_Class}
\end{figure}

\begin{figure}[ht]
    \centering
    \includegraphics[scale=0.15]{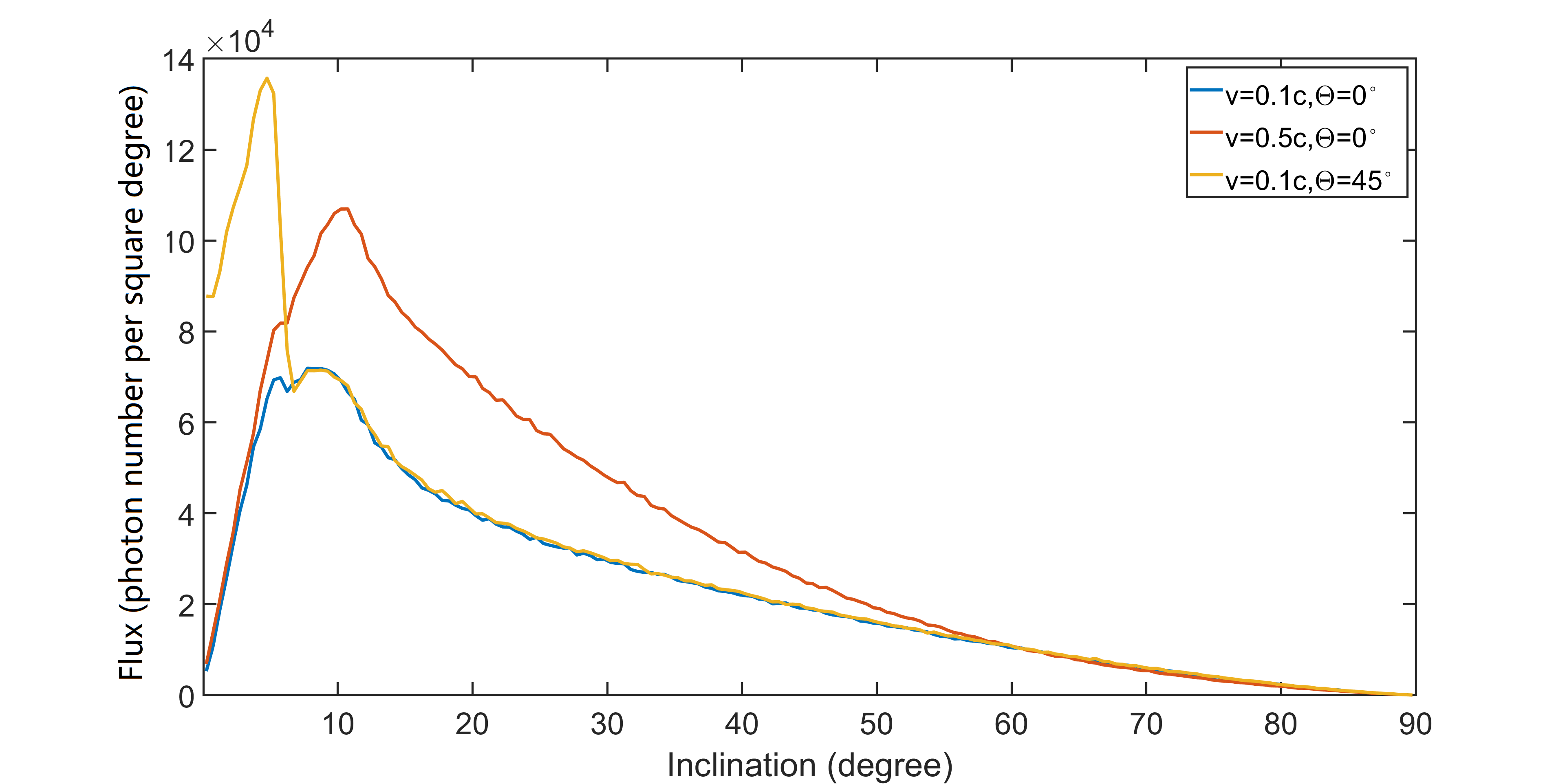}
    \caption{ \footnotesize The observed total flux of classes ``A" and ``B" photons at different observing inclination angles averaged over all azimuth angles around the $Z$ axis.}
    \label{fig:simulation_observed_flux}
\end{figure}

\begin{figure}[ht]
    \centering
    \includegraphics[scale=0.15]{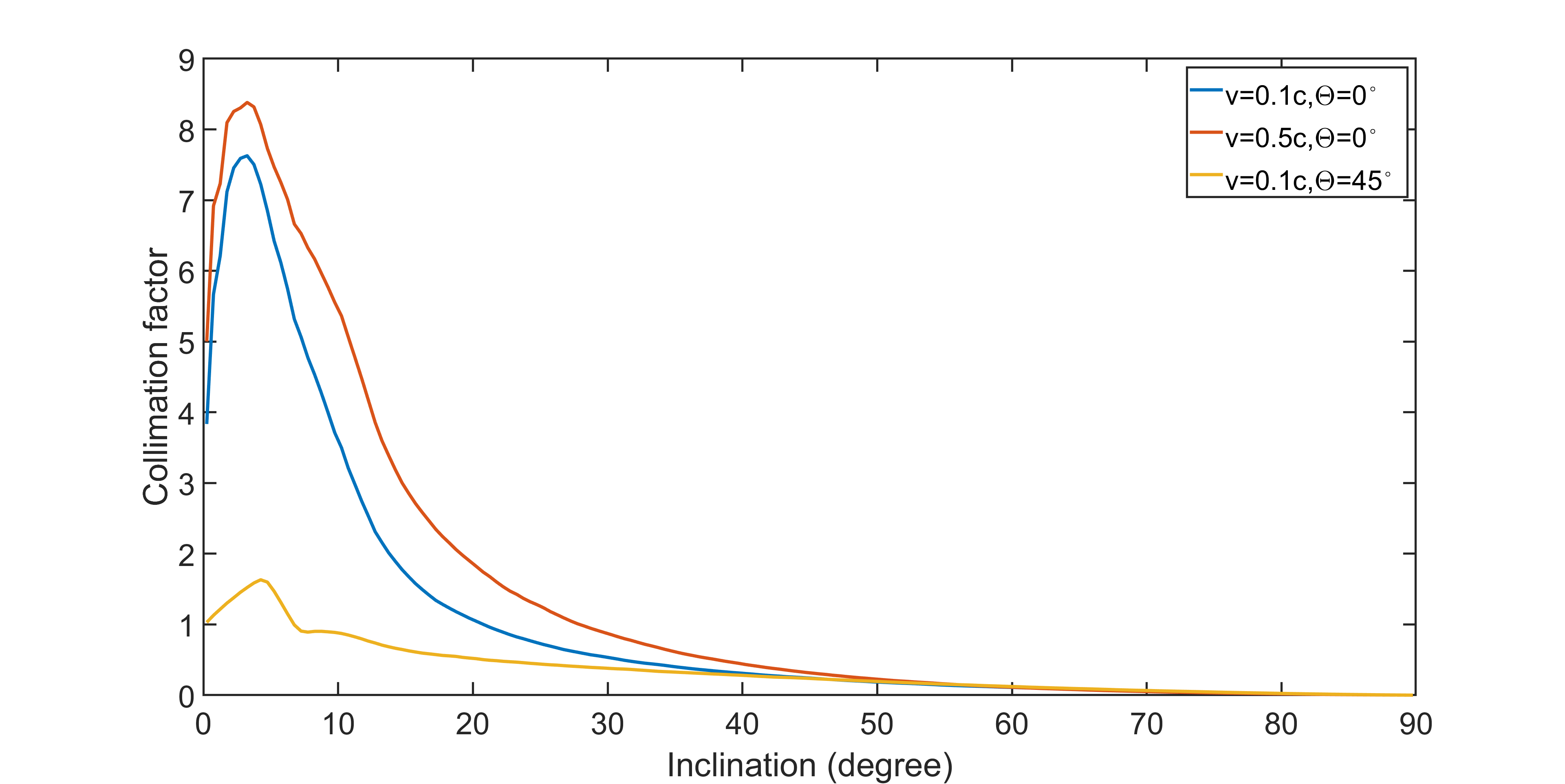}
    \caption{ \footnotesize Collimation factor as a function of observing inclination angle averaged over all azimuth angles around the $Z$ axis.}
    \label{fig:simulation_factor}
\end{figure}

\begin{figure}[ht]
    \centering
    \includegraphics[scale=0.15]{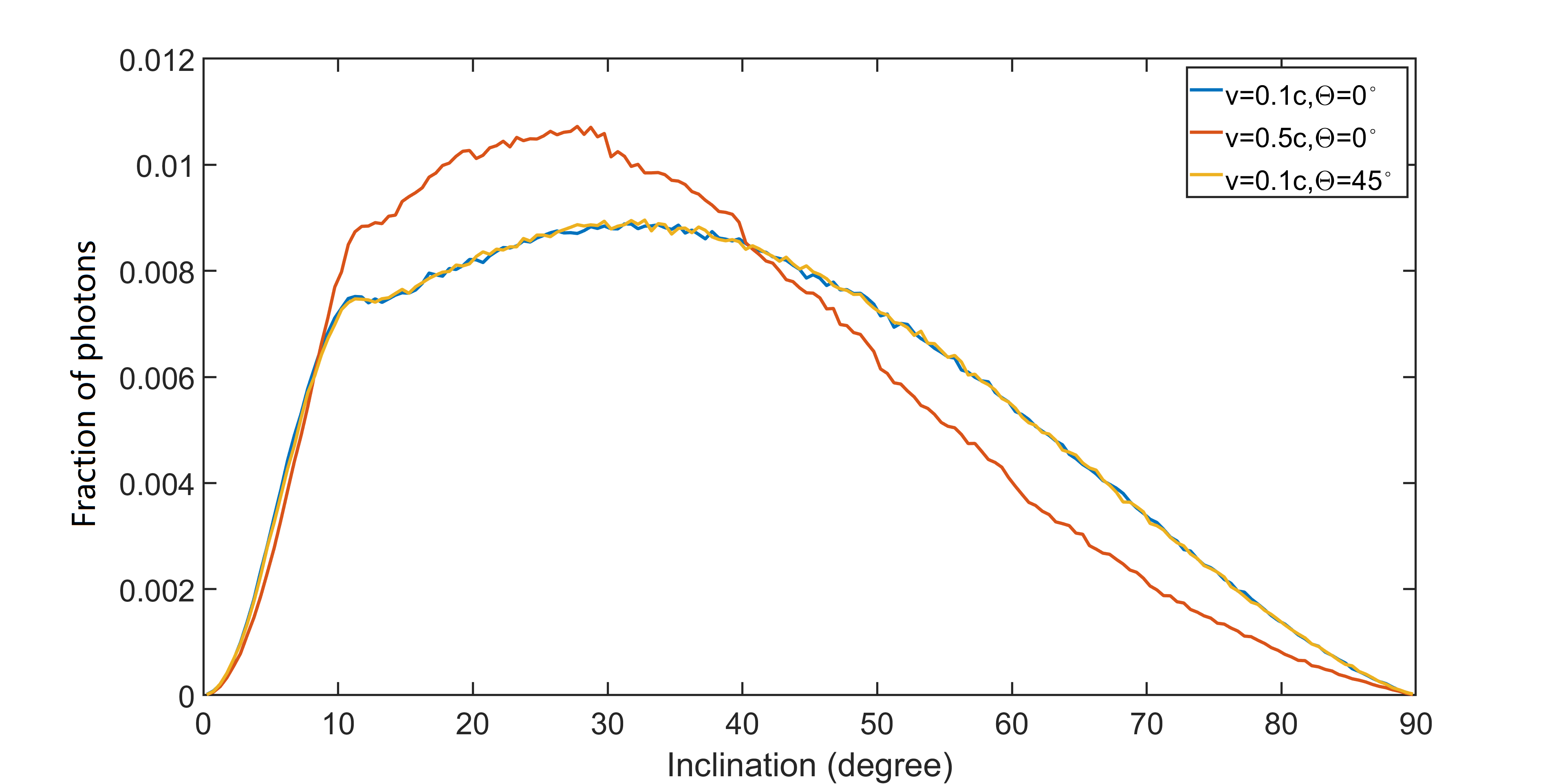}
    \caption{ \footnotesize The fraction of class ``B" photons as a functions observing inclination angle for each case of simulation. The area enclosed under each curve is normalized to unity.}
    \label{fig:simulation_inclination_B}
\end{figure}

\begin{figure}[ht]
    \centering
    \includegraphics[scale=0.15]{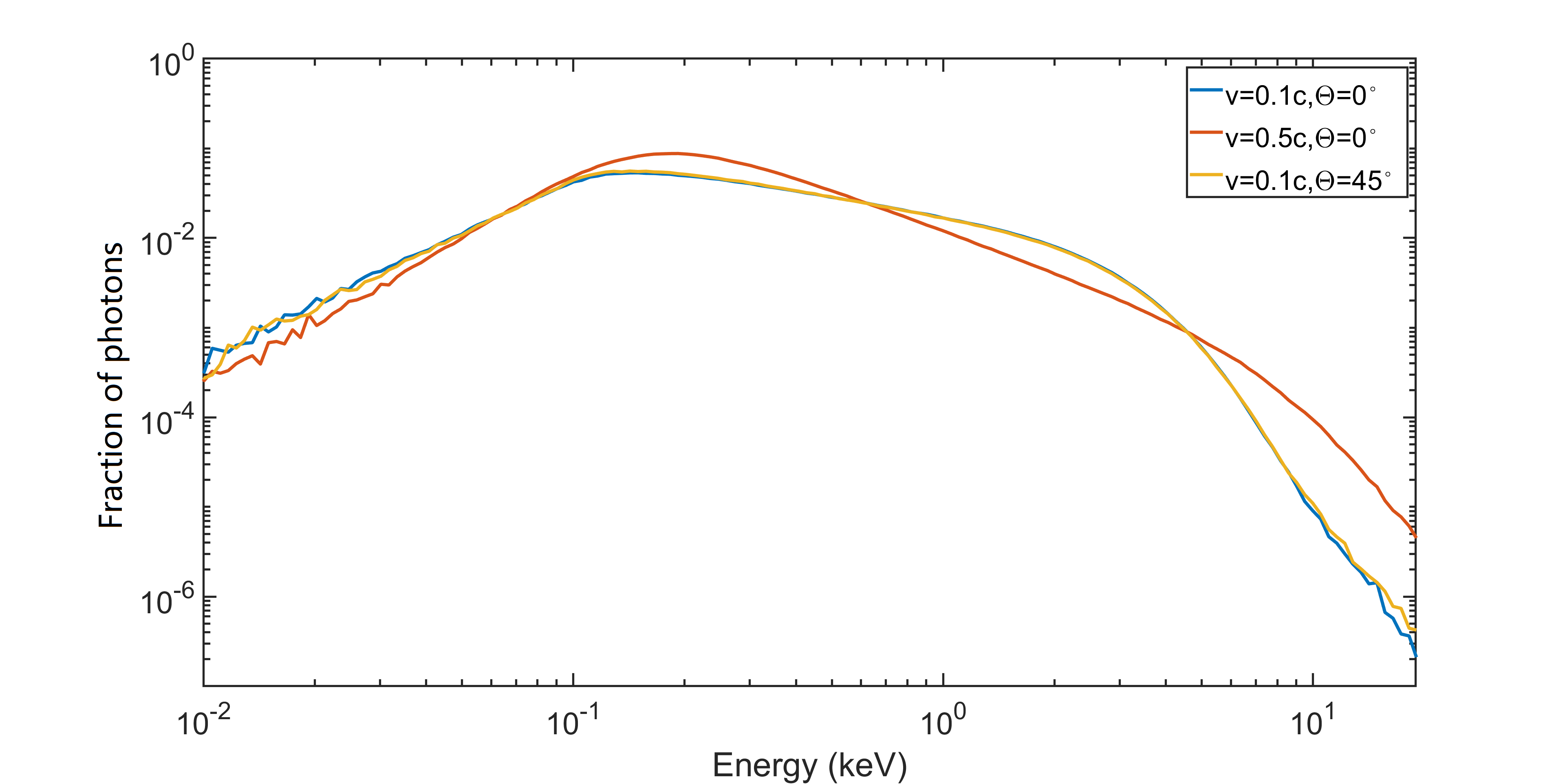}
    \caption{ \footnotesize The fraction of class ``B" photons as a function of energy for each case of simulation. The area enclosed under each curve is normalized to unity. In our simulation, the radiation of AC is totally fan beamed; if there is some pencil beamed AC radiation, the spectra of class ``B" photons will be softer.}
    \label{fig:simulation_spectra_B}
\end{figure}

\begin{figure}[ht]
    \centering
    \includegraphics[scale=0.15]{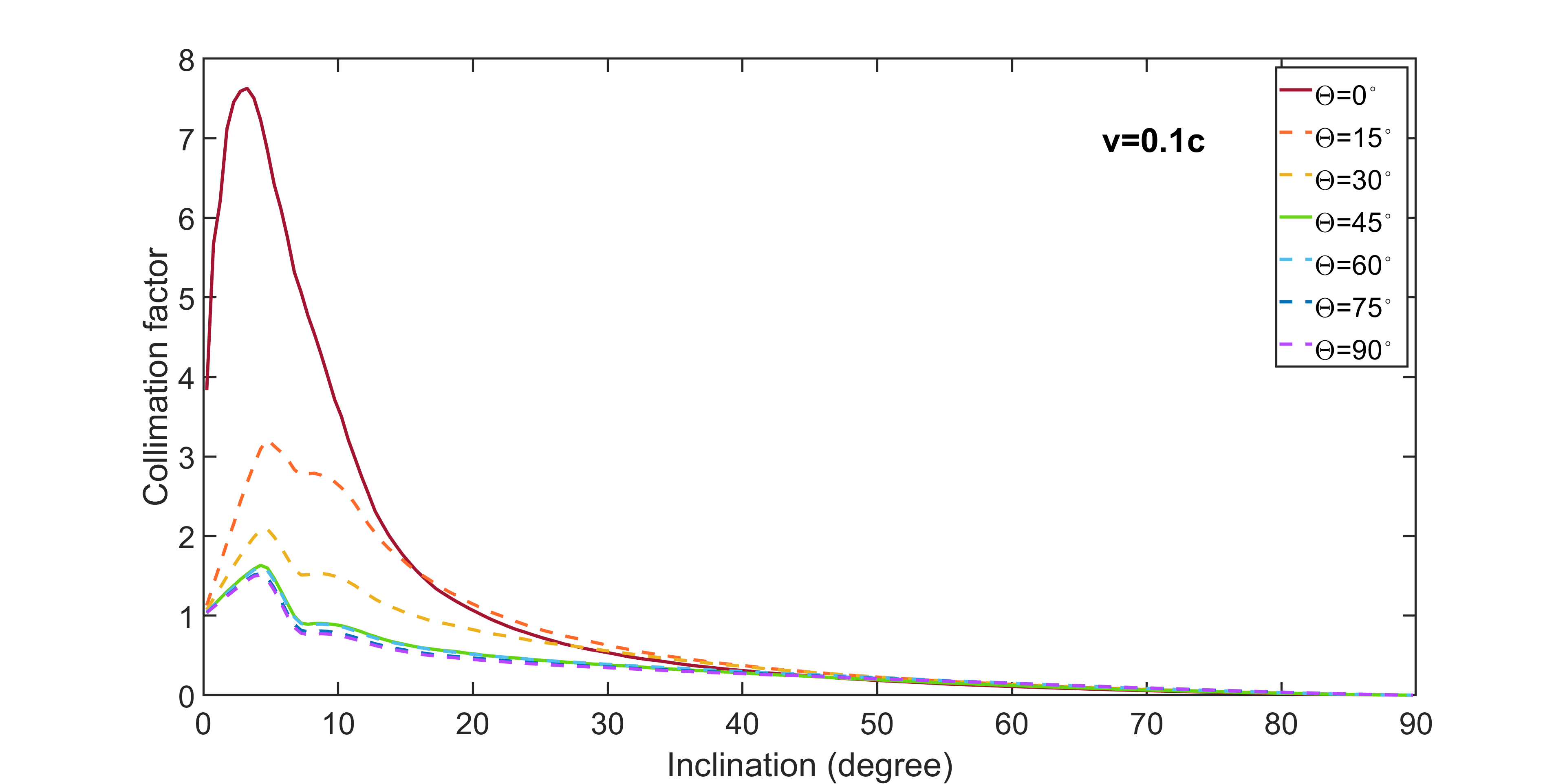}
    \caption{ \footnotesize Collimation factor as a function of observing inclination angle, for different orientation angles $\Theta$ of the AC. In all cases, $v$ is fixed to $0.1c$. The calculated collimation factors (see text for details) are drawn in dashed lines and the collimation factors directly from simulation results are drawn in solid lines.}
    \label{fig:simulation_factor_spectulation}
\end{figure}

The results of the three sets of simulations are shown in Figures \ref{fig:simulation_Class}-\ref{fig:simulation_factor_spectulation}. Figure \ref{fig:simulation_Class} is the bar chart classifying photons according to their escaping places in the system. Some photons leave the system from the wall of the cavity (``B") and more photons pass the top/bottom surfaces (``C") of the outflow. There are very few photons passing the side surface (``D") of the outflow, because photons in the outflow tend to reach the top/bottom surfaces due to the vertical bulk motions of the electrons in the outflow. Comparing the blue bars ($\Theta = 0$ and $v=0.1c$) and the red bars ($\Theta = 0$ and $v=0.5c$), it can be inferred that the flux of the cavity increases with the velocity of the outflow. However, the orientation $\Theta$ of the AC only affects strongly the percentage of photons directly leaving the cavity after being emitted from the AC (``A"). For a realistic outflow, the vertical extension can be much larger than 100 (in units of optical depth) and the ionization degree should be less for higher vertical height. The same scenario also applies to the radial direction of the outflow. Therefore, those photons escaping from the top/bottom/side surfaces of the outflow in our simulations should be absorbed in the outer part of the outflow and thus do not actually escape from the system, which are thus not considered any further.

Figure \ref{fig:simulation_observed_flux} shows the observed flux (the summed flux of classes ``A" and ``B" photons) by averaging all azimuth angles around the $Z$ axis. We define the ``collimation factor" as the ratio between the observed photon flux of the simulated outflow system and the photon flux of the AC itself without the outflow:
\begin{equation}
    f_{\rm c} = \frac{F_{\rm A}+F_{\rm B}}{F_{\rm AC}},
    \label{equation:f_collimation}
\end{equation}
where $f_{\rm c}$ is the collimation factor, $F_{\rm A}$ is the flux of class ``A" photons, $F_{\rm B}$ is the flux of class ``B" photons and $F_{\rm AC}$ is the intrinsic flux of AC without the outflow. $F_{\rm A}$ and $F_{\rm B}$ can be obtained from the simulation results, while $F_{\rm AC}$ can be analytically calculated with Eq. (\ref{equation:seed}). At small observing inclination angles, $F_{\rm A}$ is equal to $F_{\rm AC}$; when the inclination becomes larger, AC is obscured by the outflow so that $F_{\rm A}=0$. The collimation factor as a function of observing inclination angle with respect to the $Z$ axis is shown in Figure \ref{fig:simulation_factor}. It can be inferred that the velocity of outflow $v$ has a positive impact on the collimation factor, and the collimation factor decreases with $\Theta$ at small observing inclination angles. This is because the collimation factor is defined as being inversely proportional to the AC flux without outflow, which is equivalent to the flux of class ``A" photons at small observing inclination angles, and the number of class ``A" photons increases with $\Theta$. For large observing inclination angles, the collimation factor can be much less than unity, because the blocking of the outflow to the AC's direct radiation overwhelms over the scattered flux escaping from the wall of the outflow. For large values of $\Theta$, the collimation factor becomes very small.

Figures \ref{fig:simulation_inclination_B} and \ref{fig:simulation_spectra_B} show the fractions of class ``B" photons as functions of observing inclination angle and energy for different cases of simulations, respectively. From these two figures, we can find that the orientation $\Theta$ of the AC has negligible influence on the spectral shape and inclination distribution of class ``B" photons; and from Figure \ref{fig:simulation_Class} we can find that $\Theta$ has almost no influence on the total number of class ``B" photons. Therefore, we can use the flux of class ``B" photons obtained from the simulations of $\Theta=0$ or $\Theta=45^\circ$ to estimate the flux of class ``B" photons for other $\Theta$ values without performing additional simulations. We then use the flux of class ``A" and ``B" photons, and the analytically calculated flux of the AC without the outflow for different $\Theta$ values, to get the estimated collimation factor (see Eq. (\ref{equation:f_collimation})) as a function of inclination angle ($v=0.1c$), as shown in Figure \ref{fig:simulation_factor_spectulation}. The collimation factor decreases rapidly with rising $\Theta$. If the orientation of the AC and the outflow axis are independent of each other, then the collimation factor is less than 3 in most cases. 
%for other $\Theta$ values

\section{Discussion}
\label{discuss}

%\subsection{On the magnetic fields of PULXs}
We investigated the collimation effect of PULXs in the presence of outflows for the fan beam emission pattern. The above simulations are generic. First, the simulations apply to any luminosity, as far as outflows exist to collimate fan beam emission of the NS AC. Second, the NS does not rotate and thus we do not simulate the expected pulse profiles for those cases. We have carried out more extensive simulations for more realistic cases, considering different orientations of the NS's spin axis with respect to the cavity, and found that the collimated photons, i.e., ``B" photons,  have very low periodic modulation (pulse fraction $<$10\%). This means that a collimation factor $>>$ 1 will reduce the observed pulse fraction significantly (You et al. in preparation). All other results presented above remain unchanged.

Our work is an extension of that presented in \cite{Mushtukov2021} which is based on the pencil beam pattern. From our Monte-Carlo simulations, the collimation effect for fan beam emission of the AC is not strong in most cases, and can only enhance the observed flux by about one order of magnitude in rare cases (albeit with much reduced PF in conflict with the observed high PF of PULXs). In most cases, the presumed outflow actually blocks severely the intrinsic radiation of the AC. Therefore the observed sample of PULXs may be strongly biased in favor of those systems whose direct X-ray emissions from the NSs are not substantially blocked by the presumed outflows. This may be one reason for the rarity of the observed extragalactic PULXs.

\subsection{Multipole and Dipole Magnetic Fields of PULXs}

From our simulation, the luminosity of PULXs is intrinsic rather than highly amplified by geometric collimation even in the presence of strong outflows. In \cite{Mushtukov2015b}, strong magnetic fields are necessary to explain the high luminosity of PULXs and the relation between them is $L_{39}=0.35B_{12}^{3/4}$. Using the reported maximum luminosity in the literature (Table \ref{tbl-LB}), we obtain magnetic fields for the currently detected PULXs in the range of $4.6\times 10^{12}-1.8\times 10^{15}$ G, which should be considered as the lower limits of the magnetic field, since the observed luminosity is always below the allowed maximum luminosity. Note that \cite{Erkut2020} used the maximum luminosity expression which depends on the magnetic field as $311(B/B_{\rm c})^{4/3}L_{\rm E}$ \citep{Paczynski1992}, where $L_{\rm E}$ is the Eddington luminosity and $B_{\rm c}\equiv m_{\rm e}^2\,c^3/\hbar \, e = 4.4\times 10^{13}$ G is the quantum critical magnetic filed, to estimate the lower limits of the surface fields and obtained weaker values (see their Table 4) than shown above. Indeed, strong multipole magnetic fields have been proposed for the three musketeers \citep{Kong2022,Hou2022,Tsygankov2017} as well as the extragalactic PULXs NGC 5907 ULX-1 \citep{Israel2017a} and M82 X-2 \citep{Eksi2015}, in line with the theoretical calculations by \cite{Brice2021}. In these cases, the surface magnetic fields determined from the AC emission properties (e.g., critical luminosity, CRSF and maximum luminosity) are very different with (and most likely much stronger than) the dipole fields determined from torque models. However, recent studies \citep{Suleimanov2022} show that the high magnetic field increases the electron-positron pairs which compensate the opacity reduction due to the high magnetic field, such that the maximum possible luminosity does not increase with the magnetic field, contrary to what was claimed in \cite{Mushtukov2015b}.

Despite the uncertainties on the magnetic fields of PULXs from different theoretical models, it is important to answer this question: are the surface magnetic fields really much stronger than dipole fields in PULXs? The surface magnetic fields of normal accreting X-ray pulsars in the Milky Way are on the order of $10^{12}$~G, determined from the detected CRSFs shown in Table 1. Is it possible that the NSs in PULXs (including the three musketeers) with much stronger surface (multipole) magnetic fields are much younger? Most of the Galactic accreting X-ray pulsars (including the three Musketeers) are Be X-ray binaries, while extragalactic PULXs NGC 7793 P13 and NGC 300 ULX1 contain supergiant companions \citep{Motch2014,Fabrika2021}. We mention in passing that the companions in the rest of extragalactic PULXs are still unknown yet due to the lack of optical spectroscopy. Therefore, it is indeed quite possible that the NSs in extragalactic PULXs are much younger, due to their much younger companions. As these NSs becomes older, their multipole magnetic fields of $>10^{13}$~G would decay to approach $\sim 10^{12}$~G \citep{Igoshev2021}. If this is true, the stronger NS surface (multipole) magnetic fields of the three musketeers would imply that their NSs are younger than the NSs in other Galactic accreting X-ray pulsars, although it is difficult to determine their ages from their companions. Unfortunately, we currently still cannot determine accurately the dipole magnetic fields of most extragalactic PULXs, given the large uncertainties of the current torque models and limited data on the spin evolution over a larger luminosity range for extragalactic PULXs. 

One possible cause of the magnetic filed decay in accreting pulsars could be the accretion itself, such that the surface field is buried under the accreted material \citep[e.g.,][and references therein]{Ye2019}. In this scenario, persistent (extragalactic) PULXs reaching a luminosity of several $10^{39} \, \rm erg\,s^{-1}$ should have their surface magnetic fields significantly buried and thus reduced (despite of their young ages), especially compared with transient Galactic X-ray pulsars. This, however, contradicts with their very strong surface magnetic fields inferred from their very high luminosity. Therefore, accretion induced magnetic field decay unlikely plays a significant role in the magnetic field evolution of these NSs, and the stronger magnetic fields of PULXs may indeed be due to their younger ages. This would be similar to the isolated magnetars, which are quite young compared to normal NSs. For some magnetars there are large differences between their rather normal dipole magnetic fields measured from their observed spin-down rates and the much stronger magnetic fields inferred from their magnetar-unique behaviours (e.g., prolific glitches, soft gamma-ray flares, including some rare giant flares, etc.); the latter phenomena are commonly interpreted as due to the existence of strong multipole magnetic fields \citep{Ertan2007,Ertan2008,Ertan2009,Igoshev2021}.

\subsection{Outflows of PULXs}

One remaining issue is if it is reasonable to assume the existence of strong outflows for PULXs. The magnetosphere radius $R_{\rm M}$ can be calculated as \citep{Ghosh1978,Ghosh1979,Ghosh1992}
\begin{equation}
R_{\rm m} = 7\times 10^7 \Lambda\, m^{1/7}\, R_6^{10/7}\, B_{12}^{4/7}\, L_{39}^{-2/7}  \,\,\,\, \rm cm,
\end{equation}
where $\Lambda$ is a constant depending on the accretion geometry (usually taking $\Lambda=0.5$), $m$, $R_6$ and $B_{12}$ are the NS mass in units of $M_{\odot}$, radius in units of $10^6$ cm, and magnetic field in units of $10^{12}$ G, respectively. $L_{39}$ is the luminosity in units of $10^{39} \,\rm erg\,s^{-1}$. The spherization radius $R_{\rm sph}$ is \citep{Shakura1973}
\begin{equation}
R_{\rm sp} = 7.5\times10^6\,\frac{L_{39}\,R_6}{m} \,\,\,\, \rm cm.
\end{equation}
By calculating the ratio of $R_{\rm M}$ to  $R_{\rm sph}$ for different luminosities and dipole magnetic fields, we found that for a luminosity of the order of $10^{39}-10^{40}\,\rm erg\,s^{-1}$, a dipole magnetic filed of $\gtrsim 10^{13}$ makes $R_{\rm M}\gtrsim R_{\rm sph}$, i.e., the disk is not spherized outside the magnetosphere, such that there is no significant outflow driven by radiation pressure in the disk. Consequently, most of the material inflow can be channeled by the magnetic field onto the NS to form the AC, and produce super-Eddington radiation in the form of fan beam, if the surface magnetic field is strong enough to reduce the opacity, as discussed earlier. For significantly higher luminosity, strong outflow will be launched from the disk, unless the dipole magnetic fields are much stronger. 

The existence of outflows as observed or proposed for NGC 300 ULX1 \citep{Kosec2018,Vasilopoulos2019}, NGC 1313 X-2 \citep{Sathyaprakash2019,Sathyaprakash2022} and Swift J0243.6+6124 \citep{Tao2019} implies that its dipole magnetic field is not strong, but its high luminosity requires much stronger surface magnetic fields, since our simulations show that its observed luminosity should be close to its intrinsic luminosity. Therefore, co-existence of both the intrinsically ultra-high luminosity and regular dipole magnetic fields ($\sim 10^{12}$ G, comparable to most accreting X-ray pulsars) would naturally require both much stronger multipole magnetic fields ($>10^{13}$ G) and strong outflows in PULXs, including the three musketeers. However, it is possible that massive outflows may be launched if the central fan beam irradiates the disk with super-Eddington luminosity, even when $R_{\rm M}\gtrsim R_{\rm sph}$. Nevertheless, such outflows will be along the radial direction of the accretion disk and thus will not collimate the fan beam radiation.

Finally, it is reasonable to ask the question: for an accreting NS with regular dipole magnetic field (e.g. $\sim 10^{12}$ G), how is the super-Eddington luminosity (e.g., $>10^{40}\, \rm erg\ s^{-1}$) produced from the AC in the absence of significant collimation, while the disk is spherized to launch an outflow, which reduces the direct accretion onto the NS? 
%It is thus possible that a PULX with luminosity far above $10^{40}\, \rm erg\ s^{-1}$ has a regular magnetic field, i.e., $\sim 10^{13}$ G, but much higher mass transfer rate, such that the disk beyond the magnetosphere of the NS is spherized and significant outflow is produced. 
According to theoretical calculations, for a mass accretion rate of 100$-$500 (in units of Eddington rate) and $\alpha=0.1$ \citep[Eq. 2c,][]{Zhou2019}, the velocity of optically thick outflows is around 0.001$-$0.003$c$ which is about one order of magnitude lower than the escaping velocity at $R_{\rm sph}$ from where the outflow is launched. Therefore, a significant amount of outflowing material will inevitably fall back, and can be captured by the NS's magnetosphere and then channeled to form the AC, which produces the observed super-Eddington (fan beam) radiation with high PF. This scenario can be tested with numerical simulations which can follow the outflows until the expected fallback happens.

\subsection{Comparisons between Galactic and extragalactic PULXs}
It is interesting to note that, if Swift J0243.6+6124 is located in an external galaxy, like NGC 300, it will have consistent spectral parameters with extragalactic PULXs \citep{Tao2019}. In particular, the three musketeers have experienced large luminosity changes from well below their Eddington limit to super-Eddington. Their maximum luminosity fills in the luminosity gap between the Galactic X-ray pulsars and the extragalactic PULXs. This suggests a continuous distribution of sources in maximum luminosity spanning $\sim$5 orders of magnitudes. The detections of extragalactic PULXs CXOU J073709.1+653544 \citep{Trudolyubov2007} and XMMU J031747.5-663010 \citep{Trudolyubov2008} which have similar maximum luminosity as the three musketeers demonstrate that there are indeed less luminous extragalactic PULXs. 

The current Galactic PULXs are all Be X-ray binaries which become ULXs during transient outbursts, while extragalactic PULXs NGC 7793 P13 and NGC 300 ULX1 are persistent sources and contain supergiant companions \citep{Motch2014,Fabrika2021}. The question on why Galactic X-ray pulsars with supergiant companions are not ULXs can be understood easily given that among 951 galaxies which host ULXs \citep{Walton2022}, only 8 galaxies contain PULXs (Table~\ref{tbl-LB}, if including CXOU J073709.1+653544 and XMMU J031747.5$-$663010) and only the above two PULXs have been identified to have supergiant companions. Therefore only a very small percentage of accreting pulsars may have both sufficiently high mass accretion rate and reasonably strong surface magnetic field to emit at above $10^{39} \,\rm erg \,s^{-1}$ luminosity.

It is unknown whether extragalactic PULXs can be Be X-ray binaries. Nevertheless, the probability of detecting extragalactic Be PULXs has been very low, due to the transient nature of such binaries, the short peak luminosity duration, and small observational duty cycle of such transient outburst events with the current X-ray telescopes of narrow field of views. The Wide-Field X-ray (WXT) telescope on-board the future Einstein Probe (EP) mission \citep{Yuan2018} will have good sensitivity and very large field of view, thus large duty cycle in surveying for transient extragalactic PULXs. Although identification of the transient as a point source will need follow-up observations using X-ray telescopes with higher angular resolution, the Follow-up X-ray Telescope (FXT) on-board EP can be re-pointed quickly to search for pulsations, in responding to WXT triggers. We thus anticipate that EP will detect some extragalactic Be PULXs in the near future and renew our understanding of PULXs.

\acknowledgments
%This work made use of the data from the \textit{Insight}-HXMT mission, a project funded by the China National Space Administration (CNSA) and the Chinese Academy of Sciences (CAS). This research also made use of data obtained with the \textit{NICER}, a NASA experiment placed on the International Space Station (ISS). 
The \textit{Insight}-HXMT team gratefully acknowledges the support from the National Program on Key Research and Development Project (grant No. 2021YFA0718500) from the Ministry of Science and Technology of China (MOST). This work was partially supported by International Partnership Program of Chinese Academy of Sciences (Grant No.113111KYSB20190020). The authors are thankful for support from the National Natural Science Foundation of China under grants U1938103, 12041303, U1938109, U1838202, U1838201, U1838115, U1838104, 12073029, U1838107, U1938201, U2038101 and 12173103. X.H. is supported by the Light of West China Program of the CAS. L.J. is supported by the Guangdong Major Project of Basic and Applied Basic Research (Grant No. 2019B030302001).

%\appendix
%\section{X-ray pulsar luminosity and magnetic field list}

%\renewcommand{\thefootnote}{\alph{footnote}}
\begin{longrotatetable}
\begin{deluxetable*}{lcccc}
\tablecaption{Maximum luminosity and magnetic field estimation for the three musketeers, the Galactic accreting X-ray pulsars, the extragalactic ultra-luminous accreting X-ray pulsars, the LMC and the SMC X-ray pulsars, separated by horizontal lines between each group. Only sources having both maximum luminosity and magnetic field measurements are included. We do not distinguish between the dipole and multipole magnetic fields.\label{tbl-LB}}
\tablewidth{700pt}
\tabletypesize{\footnotesize}
\tablehead{
\colhead{Name\tablenotemark{a}} & \colhead{$L_{\rm max}$ (Ref)\tablenotemark{b}} & 
\colhead{$B$ (Ref)} & \colhead{$B_{\rm CRSF}$} & 
\colhead{$E_{\rm cyc}$ (Ref)}   \\ 
\colhead{} & \colhead{($\rm 10^{38} \,erg\, s^{-1}$)} & \colhead{($10^{12}\, \rm G$ )} & \colhead{($10^{12}\, \rm G$ )} & 
\colhead{($\rm keV$)}
} 
\startdata
RX J0209.6$-$7427&      	20 (1, 2, 3)&		3 (1), 28 (2), 4.8-8.6 or 1.7-2.2 (3)     \\
SMC X-3&                	25 (4)&			2.9 (5), 6.8 (6), 1-5 or 20 or 2.6 or 20-30 (4), &      \\
&                               &           		$\lesssim$3.6 (7), 0.023 (8)              \\
Swift J0243.6+6124&     	15-50 (9, 10, 11)&	10 (12, 13), $<$10 (9), 2-26 (11),   &  16&                     120-146 (14)    \\  
&				&			3-9 (15), 3.4 (16), 0.16 (8), 16 (14)&           							\\   \hline
Cen X-3&                	1 (17)&			9 or 19.1 (18)&                      			3.1&			28 (18, 19, 20)         \\
Cep X-4&                	0.06 (21, 22)&		&                                    			3.1&			28 (18, 19) 		\\       
GX 301-2&               	0.64 (23)&		0.6 or 788 (18)&                     			4.1&			37 (18, 19) 		\\     
GX 304-1&               	0.001 (24)&		0.25 (25)&                           			6.1&			54 (18, 19)  		\\     
Vela X-1&               	0.04 (26)&		172 (18)&                            			2.8&			25 (18, 19)  		\\        
X Per&                  	0.003 (27)&		9.6 or 42 (18), 40-250 (28)&         			3.3&			29 (18, 19) 		\\         
0115+63 (4U)&           	0.04 (29)&		2.8 or 5.9 (18), 0.012 (25)&         			1.3&			12 (18, 19, 30)   	\\       
0332+53 (V)&            	5 (31)&			3.4 or 4.5 (32), 3.76e-5 (25)&          		3.1&			28 (18, 19)    		\\      
0440.9+4431 (RX)&       	0.09 (33, 34, 35)&	73 (5)&                              			3.6&			32 (18, 19) 		\\         
0535+26 (A)&            	0.9 (17)&		6.6 or 296 (18), 14 (5), 0.8 (25)&        		5.6&			50 (18, 19)   		\\       
0658$-$073 (XTE)&         	0.37 (36)&		&                                    			3.7&			33 (18, 19)  		\\       
1008$-$57 (GRO)&          	0.58 (37)&		39 (5), 0.14 (25)&                   			10&			90 (37)			\\ 
1118$-$616 (1A)&          	0.29 (36)&		230 (5), 7.8 (38), 3.4 (39)&  				6.2&			55 (18, 19)		\\
1409$-$619 (MAXI)&        	1.2 (40)&		0.27 or 3.7-7 (40)&					4.9&			44 (18, 19)		\\    
1417$-$624 (2S)&          	1 (41, 42)&		$<$0.9 (43), 7 (41)											\\
1538$-$52 (4U)&           	0.07 (44)&		&							2.5&			22 (18, 19)		\\      
1553$-$542 (2S)&          	1.2 (45)&		0.52 (25)&						2.6&			23 (18, 19)		\\        
1626.6$-$5156 (Swift)&    	0.52 (36)&		0.9 (46), 3.3 (47)&					1.1&			10 (18, 19)		\\         
16393$-$4643 (IGR)&       	0.062 (48)&		&							3.3&			29 (18, 19)		\\        
16493$-$4348 (IGR)&       	0.024 (49)&		&							3.5&           		31 (18, 19)		\\         
1657$-$415 (OAO)&         	0.2 (48)&		&							4.0?&          		36? (18)		\\    
1750$-$27 (GRO)&          	1 (17, 50)&		2 (50), 4 or $<$(4-5) (51)										\\
J17544-2619 (IGR)&      	3 (52)&			&							1.9&	       		17 (18, 19)   		\\
18027$-$2016 (IGR)&       	0.28 (48)&		&							2.7&           		24 (18, 19)		\\  
18179$-$1621 (IGR)&       	0.079 (53, 54)&		&							2.4&           		21 (18, 19)		\\     
1829$-$098 (XTE)&         	0.043 (55)&		&							1.7&           		15 (18, 19)		\\       
1843+009 (GS)&          	0.3 (56)&               &							2.2?           		20? (18)		\\       
1858+034 (XTE)&         	0.2 (57, 58)&           &							5.4&           		48 (57, 58)		\\     
1901+03 (4U)&           	1.2 (48)&               0.3 or $<$0.5 (59), 0.31 (60), 4.3 (61)&		1.1?, 3.4?&    		10? (18), 30? (62)	\\
1907+09 (4U)&           	0.23 (27)&              10 (63)&						2.0&           		18 (18, 19)		\\
1908+075 (4U)&          	0.084 (48)&             $>$10 (64)&						4.9?&          		44? (18)		\\
19294+1816 (IGR)&       	0.001 (65)&             &							4.8&           		43 (18, 19)		\\
1946+274 (XTE)&         	0.8 (66)&               1.23 (25)&						4.0&           		36 (18, 19)		\\
1947+300 (KS)&          	1.3 (67)&               1.2 (67), 5.33 (25), 25 or 16 or 30 (68)&		1.3&           		12 (18, 19)		\\
2030+375 (EXO)&         	2.7 (69)&               4-6 (69), 1-4 (70)&					1.2? 4.0/7.1?&  	11? (71), 36/63? (18)	\\
2058+42 (GRO)&          	0.58 (25)&              1-2 (72), 37 (25)&					1.1&           		10 (72, 73)		\\
2103.5+4545 (SAX)&      	0.084 (48)&             1 (74), 16.5 (75), $<$5.2 (76), 16-30 (77), 12 (78)&	1.3?&     		12? (18)		\\
2206+54 (4U)&           	0.04 (79)&              $>$20 (80), 4 (81), 60-90 (82)&				3.3-3.9?&      		29-35? (18)		\\   \hline
M51 ULX-7&             		70 (83, 84)&     	1-10 (83), 20-70 (84), 0.0069 or 0.19 (8),&								\\
&                     		&			0.62-1 or 0.62-60 or 7-260 (85)&									\\  
M82 X-2&               		180 (86)&         1 or 70 or 100 or $\gtrsim$100 or $\lesssim$0.001 (87), 							\\ 
&                               &                 0.09 (8), 2-2.9 or 2-50 or 3-100 (85), 5-20 (88)&              						\\  
NGC 300 ULX1&          		47 (89)&          	3 or 10 or 20 (89), 6 or 10 (90), 1.2 (8)&          	1.5&   			13 (18)			\\
&				&			6.8 or 6.8-2400 or 120-3500 (85)&									\\         
NGC 1313 X-2&          		200 (91)&        	1.44 (92), 40-250 or 3.5-210 (85)&									\\   
NGC 5907 ULX-1&			1000 (93)&       	184 (92), $>$(70-300) (93), 21 (8),&		       							\\    
&				&			27-38 or 56-94 or 27-250 or 2.4-450 (85)&								\\
NGC 7793 P13&			100 (94, 95)&    	4.17 (92), 1.5 (94), $>$80 (95), 0.25 (8),&								\\  
&				&			0.12-0.32 or 2.2-12 or 0.12-5 or 0.77-33 (85)								\\  
CXOU J073709.1+653544&          		12 (96)   &            	0.56 (8) 												\\  \hline
LXP4.10&			0.37 (97)&		$>$0.3 (97)												\\    
LXP4.40&			0.08 (97)&		$>$0.3 (97)												\\ 
LXP8.03&			8 (56)&			0.06 (25), $>$0.3 (97)&					3.5&           		31 (18, 19)		\\ 
LXP13.5 (LMC X-4)&		3.6 (98)&		27-32 (98)&						11.2?&         		100? (18)		\\ 
LXP27.2&			1.3 (99)&		10 (99), 15 (100)											\\ 
LXP38.55&			0.03 (101)&		0.01 (101)												\\  
LXP61.6&			3 (102)&		&							1.1?/2.2?&     		10?/20? (18, 19)	\\ 
LXP187&				0.07 (103)&		120 (103)&												\\    \hline
SXP0.72 (SMC X-1)&		5 (104)&		&							5.6&           		50 (105)		\\    
SXP2.16&			0.0069 (106)&		0.1-0.4 (107)												\\ 
SXP2.37 (SMC X-2)&		5.5 (108)&		3 (109)&						3&             		27 (18, 19)		\\     	   
SXP2.76&			1 (110)&		&							          					\\      
SXP4.78&			2.5 (111)&		$<$1.5 or $<1$ or 0.6 or 2 (111)&			1.1?&          		10.2? (18)		\\    
SXP6.85&			0.33 (27)&		2.1 (5)													\\ 
SXP8.80&			0.73 (27)&		4.1 (5)													\\ 
SXP15.3&			1.2 (112)&		6 (5), $<$15 (7)&					0.6&           5(18) 				\\         
SXP15.6&			0.083 (113)&		$<$5.2 (113)									 			\\          
SXP18.3&			0.3 (27)&		5 (5)													\\        
SXP25.5&			0.036 (5)&		6 (5)													\\        
SXP46.6&			0.074 (56)&		12 (5), $<$6 or $<$4 (7)										\\       
SXP59.0&			1 (114)&		1-10 (114), 23 (5), $<$14 or $<$13 (7)						 			\\          
SXP74.7&			0.35 (27)&		38 (5)													\\
SXP82.4&			0.05 (5)&		27 (5)													\\
SXP91.1&			0.29 (56)&		19 (5), $<$25 (7)					 						\\
SXP95.2&			0.2 (17)&		38 (5)													\\
SXP101&				0.033 (5)&		27 (5)													\\
SXP140&				0.04 (5)&		43 (5)							 						\\
SXP144&				0.011 (7)&		$<$24 (7)						 						\\ 
SXP152&				0.039 (5)&		51 (5)													\\
SXP169&				0.2 (56)&		71 (5), $<$30 or $<$26 (7)										\\
SXP172&				0.056 (56)&		56 (5)							 						\\
SXP175&				0.05 (5)&		71 (5)													\\
SXP202A&			0.083 (106)&		78 (5)													\\
SXP202B&			0.045 (115)&		64 (5)													\\
SXP214&				0.029 (5)&		68 (5)													\\
SXP264&				0.021 (5)&		72 (5)													\\
SXP280&				0.029 (5)&		85 (5)													\\
SXP293&				0.028 (5)&		93 (5)													\\
SXP304&				0.068 (5)&		148 (5)													\\
SXP323&				0.055 (5)&		134 (5)													\\
SXP327&				0.02 (116)&		69 (5)													\\
SXP342&				0.043 (5)&		146 (5)													\\               
SXP343&				0.015 (16)&		4-12 (117)												\\
SXP455&				0.07 (5)&		248 (5)													\\
SXP504&				0.035 (5)&		196 (5)													\\
SXP565&				0.019 (5)&		161 (5)													\\
SXP645&				0.025 (5)&		228 (5)													\\
SXP701&				0.027 (5)&		250 (5)													\\
SXP726&				0.048 (5)&		404 (5)													\\
SXP756&				0.063 (5)&		419 (5), $<$270 or $<$3 or $<$200 (118)									\\
SXP893&				0.024 (5)&		310 (5)													\\
SXP967&				0.071 (5)&		595 (5)													\\
SXP1062&			0.0083 (106)&		10 (119), 3 or 7 (120), $\gtrsim$100 (121)								\\
&				&			40 (122), 150 (123)									        \\
SXP1323&			0.097 (5)&		996 (5)								                  \\			
\enddata
 % footnote:
\tablenotetext{a}{The list of Galactic accreting X-ray pulsars is a combination of \cite{Staubert2019}, \cite{Malacaria2020} and \cite{Karino2007}, but with LMXB, SGR and AXP excluded. The LMC X-ray binary pulsar list is from http://www.southampton.ac.uk/$~$mjcoe. The SMC X-ray binary pulsar list is a combination of \cite{Klus2014} and \cite{Yang2017}.}
\tablenotetext{b}{$L_{\rm max}$ denotes the maximum luminosity found in the literature. It is the peak luminosity during Type II outbursts if available, or else it is the average or maximum luminosity from multiple observations and/or from single one which could be also during a Type II outburst. Type I outbursts and flares are excluded.}
\tablerefs{
1: \cite{Vasilopoulos20201}; 2: \cite{Chandra2020}; 3: \cite{Hou2022}; 4: \cite{Tsygankov2017}; 5: \cite{Klus2014}; 6: \cite{Weng2017}; 7: \cite{Galache2008}; 8: \cite{King2020}; 9: \cite{Tsygankov2018}; 10: \cite{van2018}; 11: \cite{Wilson-Hodge2018}; 12: \cite{Doroshenko2018}; 13: \cite{Zhang2019}; 14: \cite{Kong2022}; 15: \cite{Doroshenko2020}; 16: \cite{Sugizaki2020}; 17: \cite{Karino2007}; 18: \cite{Staubert2019}; 19: \cite{Ye2019}; 20: \cite{Tomar2021}; 21: \cite{Furst2015}; 22: \cite{Vybornov2017}; 23: \cite{Liu2018}; 24: \cite{Escorial2018}; 25: \cite{Kabiraj2020}; 26: \cite{Lomaeva2020}; 27: \cite{Brown2018}; 28: \cite{Yatabe2018}; 29: \cite{Wang2011}; 30: \cite{Bissinger2020}; 31: \cite{Tsygankov2006}; 32: \cite{Doroshenko2017}; 33: \cite{Usui2012}; 34: \cite{Tsygankov2012}; 35: \cite{Ferrigno2013}; 36: \cite{Reig2013}; 37: \cite{Ge2020}; 38: \cite{Nespoli2011}; 39: \cite{Devasia2011}; 40: \cite{Donmez2020}; 41: \cite{Ji2020}; 42: \cite{Gupta2019}; 43: \cite{Inam2004}; 44: \cite{Hemphill2016}; 45: \cite{Lutovinov2016}; 46: \cite{Icdem2011}; 47: \cite{DeCesar2013}; 48: \cite{Sidoli2018}; 49: \cite{Coley2019}; 50: \cite{Shaw2009}; 51: \cite{Lutovinov2019}; 52: \cite{Romano2015}; 53: \cite{Nowak2012}; 54: \cite{Bozzo2012}; 55: \cite{Shtykovsky2019}; 56: \cite{Raguzova2005}; 57: \cite{Malacaria2021}; 58: \cite{Tsygankov2021}; 59: \cite{Galloway2005}; 60: \cite{James2011}; 61: \cite{Tuo2020}; 62: \cite{Beri2021}; 63: \cite{Doroshenko2012}; 64: \cite{Jaisawal2020}; 65: \cite{Tsygankov2019}; 66: \cite{Wilson2003}; 67: \cite{Epili2016}; 68: \cite{Tsygankov2005}; 69: \cite{Epili2017}; 70: \cite{Klochkov2007}; 71: \cite{Wilson2008}; 72: \cite{Molkov2019}; 73: \cite{Mukerjee2020}; 74: \cite{Sidoli2005}; 75: \cite{Baykal2007}; 76: \cite{Reig2014}; 77: \cite{Ducci2008}; 78: \cite{Baykal2002}; 79: \cite{Wang2010}; 80: \cite{Torrejon2018}; 81: \cite{Ikhsanov2013}; 82: \cite{Ikhsanov2010}; 83: \cite{Castillo2020}; 84: \cite{Vasilopoulos20202}; 85: \cite{Erkut2020}; 86: \cite{Bachetti2014}; 87: \cite{Kaaret2017}; 88: \cite{DallOsso2015}; 89: \cite{Carpano2018}; 90: \cite{Vasilopoulos2018a}; 91: \cite{Sathyaprakash2019}; 92: \citep{Koliopanos2017}; 93: \cite{Israel2017a}; 94: \cite{Furst2016}; 95: \cite{Israel2017b}; 96: \cite{Trudolyubov2007}; 97: \cite{Christodoulou2016}; 98: \cite{Shtykovsky2017}; 99: \cite{Coe2015b}; 100: \cite{Sahiner2016}; 101: \cite{Vasilopoulos2016}; 102: \cite{Manousakis2009}; 103: \cite{Klus2013}; 104: \cite{Hu2019}; 105: \cite{Pradhan2020}; 106: \cite{Kennea2018}; 107: \cite{Vasilopoulos2017a}; 108: \cite{Jaisawal2016}; 109: \cite{Lutovinov2017}; 110: \cite{Kohno2000}; 111: \cite{Semena2019}; 112: \cite{Maitra2018}; 113: \cite{Vasilopoulos2017b}; 114: \cite{Weng2019}; 115: \cite{Haberl2008b}; 116: \cite{Coe2008}; 117: \cite{Israel2000}; 118: \cite{Ducci2018}; 119: \cite{Popov2012}; 120: \cite{Gonzalez2018}; 121: \cite{Fu2012}; 122: \cite{Ikhsanov2012}; 123: \cite{Serim2017}; 124: \cite{Pozdnyakov1983}
}
\end{deluxetable*}
\end{longrotatetable}

\bibliography{RXJ0209.bib}{}
\bibliographystyle{aasjournal}

\end{document}